# Enhanced Flux Pinning in $YBa_2Cu_3O_{7-\delta}$ Films by Nano-Scaled Substrate Surface Roughness

Zu-Xin Ye,[a] W. D. Si,[b] Qiang Li,[a] Y. Hu,[a] P. D. Johnson,[b] and Y. Zhu[c]

[a]Materials Science Department, Brookhaven National Laboratory, Upton, New York 11973

[b]Physics Department, Brookhaven National Laboratory, Upton, NY 11973

[c]Center for Functional Nanomaterials, Brookhaven National Laboratory, Upton, NY 11973

Nano-scaled substrate surface roughness is shown to strongly influence the critical current density ($J_c$) in $YBa_2Cu_3O_{7-\delta}$ (YBCO) films made by pulsed-laser deposition on the crystalline $LaAlO_3$ substrates consisting of two separate twin-free and twin-rich regions. The nano-scaled corrugated substrate surface was created in the twin-rich region during the deposition process. Using magneto-optical imaging techniques coupled with optical and atomic force microscopy (AFM), we observed an enhanced flux pinning in the YBCO films in the twin-rich region, resulted in ~ 30% increase in $J_c$, which was unambiguously confirmed by the direct transport measurement.

Correspondence should be addressed to qiangli@bnl.gov

In large-scale applications of high $T_c$ superconductors (HTS), the key candidate is the so-called second generation coated conductors, where $YBa_2Cu_3O_{7-\delta}$ (YBCO) films are deposited on buffered metal substrates bi-axially aligned.[1] Two of the leading factors for limiting $J_c$ in YBCO coated conductors are the degree of texture and surface roughness of the buffered metal substrate. The former stems from the grain boundary problem, i.e. boundaries between misaligned YBCO grains carry significantly lower $J_c$ than the grains. This has been thoroughly studied using YBCO films grown on $SrTiO_3$ (STO) bi-crystal substrate.[2] Recently, Foltyn and coworkers achieved very high $J_c$ in YBCO coated conductors on electropolished Hastelloy C276 substrate.[3] However, it was not clear whether the enhancement was due to the smoother substrate surface or to the improved in-plane texture of the buffer layer. The effect of substrate surface roughness on $J_c$ in YBCO films has not previously been explored in isolation from other effects, like substrate texture and interlayer diffusion. This is probably due to the difficulty in preparing single crystal substrates with controllable nano-scaled surface roughness, comparable to the metal substrate used in coated conductors, which are, at the same time, suitable for growth of high quality YBCO films.

The goal of this work is to perform a study where the influence of substrate surface roughness on the $J_c$ in YBCO films can be isolated. We took advantage of the surface reconstruction of twined $LaAlO_3$ (LAO) single crystal substrates to produce a surface roughness of 2-6 nm typically found on the surface of metal substrates used for YBCO coated conductors. Unlike STO or MgO substrates, LAO crystals are usually heavily twinned due to its orthorhombicity. Among commercially available large LAO crystal wafers (~ 5 cm diameter), it is sometimes possible to find small twin-free regions. Hence, a substrate can be selected which consists of two regions, one twinned and one twin-free.

The equilibrium high temperature cubic (HTC) perovskite to low temperature orthorhombic (LTO) transition temperature of LAO is ~ 544°C.[4] However, there is significant hysteresis. In fact, even when orthorhombic substrates are heated to a typical deposition temperature of 800°C, the orthorhombic structure is retained for a while. At the deposition temperature, the orthorhombic surface reconstructs, which results in a corrugated surface with a relief depth of ~ 4.5 nm.[4, 5] In Figure 1, we illustrated the process we utilized in a schematic drawing of *LTO twinned area* of a selected LAO single crystal substrate having both twinned and twin-free regions. Figure 1(a) illustrates the first step, where the LAO (100) surface is polished to sub-nm smoothness. In the next step the substrate is brought to pulsed laser deposition (PLD) temperature (790°C). The LAO does not transform to HTC during the deposition time, but the heating does allow surface reconstruction of the (100) LTO surface prior to deposition, to provide the desired surface roughness of ~ 4.5 nm. Deposition of YBCO takes place onto the corrugated LAO (Fig. 1b). Note that at the same time, on another portion of the substrate, deposition takes place on the flat twin-free region. This allows a direct comparison of YBCO films deposited under identical conditions onto the corrugated twinned and smooth twin-free LAO substrate. Figure 1(c) shows the corrugated interface between YBCO and twinned LAO substrate when sample is cooled to room temperature, where possible twin motion is also illustrated. Specifically, in this study the c-axis oriented 0.2 μm thick YBCO films were grown using a KrF excimer laser on such selected (100) LAO substrates polished to sub-nanometer smoothness. The substrates were heated to 790°C, and the 5 minutes deposition took place in an oxygen environment of 100 mTorr. Subsequently, samples were brought to room temperature at ~ 60°C per minute. The films were of good quality, with superconducting transition temperature $T_c$ =89.2 K and $\Delta T_c$ ~1.5 K. Transport $J_c$ (~ $6\times10^6$ A/cm$^2$) at 77 K is comparable to that of the best YBCO films deposited on STO.[6]

Magneto-optical imaging (MOI) studies were performed using a low temperature MOI station described previously.[7] A high-resolution garnet MOI indicator film was placed directly onto the sample surface, with magnetic field $B_a$ always applied perpendicular to the film surface, and polarizer and analyser crossed at 90 degrees. Brightness intensity in an MOI image is proportional to the local magnetic field normal to the surface of the film.[7, 8] To coordinate the MOI studies, both optical and atomic force microscopy (AFM) were used to examine the surface of the YBCO films, as well as the underlying substrate surface after removal of the YBCO films. Transmission electron microscopy (TEM) was used to evaluate the microstructures at YBCO-LAO interface.

Figure 2 shows the corresponding images of one 0.2-μm-thick YBCO film patterned into an 0.8 mm wide strip, (a) MOI, (b) (c) optical microscopy, and (d) (e) AFM. MOI shown in Figure 2(a) was taken at 30 K and $B_a$ = 50 mT applied to the zero-field-cooled (ZFC) sample. Small notches at the edge of the films indicated by the arrows were due to fabrication/handling of the films, and did not produce any significant change in the overall pattern.[7, 8]   In general, a uniform flux penetration was observed and consistent with the prediction of the critical state model for a type II superconducting film.[7, 8]

Figures 2(b) and 2(c) are the optical images of the film taken at left and right ends, respectively. The twin-free region was at the far left end of the YBCO strip, while the twin-rich region occupied the rest of the substrate which included the right end of the film as shown in Figure 2(c). Magnetic flux clearly penetrated deeper into the sample at the left twin free region than that at the right twin rich region, suggesting a different flux pinning strength. The LAO substrate twin domains with a width of 30-100 $\mu$m can be seen through the 0.2-$\mu$m-thick YBCO film, as well as on the area where the YBCO films were etched off.  It should be emphasized that the twin structures shown in Figures 2(b) and 2(c) are the twin domains of LAO single crystals, not in YBCO films.  It is known that PLD-processed YBCO films have small growth islands due to the screw dislocations, which results in micro-twins usually crossed 90$^o$.[9]  Our subsequent TEM examination of the YBCO-LAO interface confirmed that there was no correlation between the twining in the YBCO film and that in the LAO substrate.

The dotted areas shown in Figure 2(b) and 2(c) were free of processing defects, and hence used for direct transport measurements of $J_c$ after MOI and AFM analysis. In the twin-free area, a 34 $\mu$m wide bridge, shown at the bottom of Figure 2(b), was patterned by focused laser beam. In the twin-rich area, two bridges, shown at the bottom of Figure 2(c), were patterned with one (31 $\mu$m wide, horizontal) parallel to the substrate twin domain boundaries, while the other (36 $\mu$m wide, vertical) perpendicular to the twin domain boundaries.

Figures 2(d) and 2(e) are the AFM images of the film in the twin-free and twin-rich regions. In the twin-free region, the averaged size of the YBCO growth island is ~ 0.3 μm with mean surface roughness ~ 5 nm, while in the twin-rich region, it is ~ 0.7 μm with mean surface roughness ~ 9 nm. It is important to note that the excessive surface roughness on the films in the twin-rich region (~ 4 nm) is consistent with the expected relief depth of ~ 4.5 nm due to the surface reconstruction there.

To determine the local $J_c$ in the YBCO films at both the twin-free and the twin–rich regions, we performed a detailed analysis of the MOI images taken at various temperatures and field. For simplicity, we employed the Bean critical state model with a field-independent $J_c$ for a type II superconducting narrow strip of width $w$ and thickness $d$ under perpendicular $B_a$. $J_c$ is determined by measuring the position of the flux front in a ZFC superconducting film, via the formula, $2a/w = 1/cosh(B_a/B_d)$, where $a$ is a distance between the flux front and the center line of the strip, and $B_d = \mu_0 J_c d/2$. A detailed description of this formula can be found in Ref. 7, 8 and 10. At 77 K, corresponding values of $J_c$ at twin-free ($J_c^{\text{f,M}}$) and twin-rich ($J_c^{\text{r,M}}$) areas were $4.2 \times 10^6$ A/cm$^2$ and $6.2 \times 10^6$ A/cm$^2$. It must be emphasized that the flux front position (or the value of $a$) at relatively high field, such as 50 mT used in Fig. 2(a), are affected by the flux front bending at the corner of the film and notches at the film edge. However, at low fields (e.g 15 mT at 30 K), the flux penetration depth is small so that a large portion of flat flux front can be found free of the influence of corner bending and notches, which take only a very small portion of the flux front.

For unambiguous confirmation, we performed direct transport $J_c$ measurements of each micro-bridge at liquid nitrogen temperature using 0.5 ms (settle time) pulse current to avoid sample heating. We obtained direct transport $J_c^{\text{f,T}} = 3.7 \times 10^6$ A/cm$^2$ at 77 K in the twin-free region, whereas in the twin-rich region, $J_c^{\text{r,T,Para}} = 6.4 \times 10^6$ A/cm$^2$ for the bridge parallel to the twin domain boundaries, and $J_c^{\text{r,T,Perp}} = 6.7 \times 10^6$ A/cm$^2$ for the bridge perpendicular to the twin domain boundaries. A table inset to Figure 3 listed all $J_c$ values determined at 77 K, clearly showing consistent $J_c$ values determined by MOI and direct transport methods. It is quite remarkable that the corrugated substrate surface enhances flux pinning, and results in an increase of $J_c$ in YBCO film of over 30 % at 77 K. Furthermore, both MOI and transport measurements showed negligible $J_c$ anisotropy in the twin-rich region with respect to current flowing directions parallel or perpendicular to the LAO substrate domain boundaries.

Figure 3 shows the relative $J_c$ enhancement, expressed as $\Delta J_c / \langle J_c \rangle \equiv 2(J_c^{\text{r}} - J_c^{\text{f}})/(J_c^{\text{r}} + J_c^{\text{f}})$, as a function of $T$. $\Delta J_c / \langle J_c \rangle$ is in general > 25% and slowly increases as $T$ increases. The detailed investigation of the origin of the pinning by the corrugated substrate surface is beyond the scope of the present work. Surface pinning due to the variation of vortex line energy is typically cited as a source of enhanced pinning related to the corrugated surface. However, the different surface morphologies observed from the AFM studies at the twin-free and twin-rich regions could suggest that this corrugation might have induced additional structural defects during the growth of YBCO films and that contributed more pinning. In the epitaxial YBCO thin films grown on vicinal STO (001) surface, Jooss and coworkers reported enhanced vortex pinning due to strong quasiparticle scattering at an antiphase boundary.[11]

Finally, it should be emphasized that our observation of $J_c$ enhancement in YBCO films by the nano-scaled substrate surface roughness is fundamentally different from the earlier MOI studies on the effect of twin boundaries *inside* the YBCO single crystals on flux motion. In the unidirectional twinned YBCO single crystals, several earlier studies reported a preferential flux penetration along the twin boundary (or anisotropic flux pinning).[12] Unlike YBCO single crystals, our PLD-processed YBCO films have micro-twins in each growth island, generally crossed $90^o$ throughout the film, and hence completely independent of the twin domain structures observed in the underlying LAO substrates. AFM images showed no observable difference in the film surface morphology parallel or perpendicular to the twin domains in the LAO substrate. Both direct transport measurements and MOI images, showing flux penetration length is the same along or across the twin domain boundaries in LAO, clearly indicate the feature of an isotropic pinning force. This observation is perhaps different from our perception that the corrugation might introduce directional structural defects in YBCO films, e.g. antiphase boundaries, and produce $J_c$ anisotropy. Nonetheless, our observation is consistent with numerous reports on PLD grown YBCO films on twined LAO substrate, where $J_c$ anisotropy has never been found.

In summary, an coordinated analysis of MOI, direct transport, optical microscopy, AFM and TEM in YBCO film grown by PLD on LAO substrate revealed a remarkable 30% enhancement of $J_c$ due to nano-scaled substrate surface roughness. This unambiguous finding in the case of PLD-processed films suggests that some level of nano-scaled surface roughness on the buffered metal substrate might in fact help to improve the flux pinning capability in coated conductors, provided that the overlaying YBCO films can be grown satisfactorily.

ACKNOWLEDGMENT

The authors would like to thank L. Davis for helping in the AFM characterization. The work was supported by the U. S. Dept. of Energy, Office of Basic Energy Science, under contract No. DE-AC-02-98CH10886.

FIGURE CAPTIONS

Fig. 1  Schematic drawing illustrating the formation of a nano-scaled corrugated interface between YBCO and twinned LAO substrate.  (a) (100) cut and polished LAO at room temperature with natural twin domains; (b) Initial deposition of YBCO films on LAO at ~ 790$^{\mathrm{o}}$C, where (100) surface reconstruction at high temperature leads to a corrugation on LAO surface prior to the deposition. (c) The corrugated interface between YBCO and LAO when sample is cooled to room temperature, where possible twin domain motion in LAO is also shown.

Fig. 2  The corresponding magneto-optical image (MOI), optical microscopy and AFM taken on the same 0.2-$\mu$m-thick YBCO film strip (0.8 mm wide). (a) MOI was taken at 30 K and external field $B_{\mathrm{a}}$ = 50 mT applied to the ZFC sample. Small pinholes or notches at the edge of the films indicated by the arrows were due to fabrication/handling of the films. (b), (c) Optical images, taken at left and right ends of the film strip, respectively, show the twin-free region (b) at the far left end of the strip, while the twin-rich region occupied the rest of the substrate which included the right end of the strip (c). In the dotted areas, micro-bridges (shown in enlarged view at bottom) were patterned for direct transport $J_{\mathrm{c}}$ measurement, after MOI and AFM analysis. (d), (e) AFM images of the film in the twin-free and the twin-rich regions, respectively.  In the twin-free region (d), the averaged YBCO growth island is ~ 0.3 $\mu$m with mean

surface roughness ~ 5 nm, while in the twin-rich region, it is ~ 0.7 $\mu$m with mean surface roughness ~ 9 nm.

Fig. 3    Temperature dependence of MOI determined $J_c$ in the twin-free ($J_c^{\text{f}}$, solid symbols) and the twin-rich ($J_c^{\text{r}}$, open symbols) regions. The inset figure shows the relative $J_c$ enhancement at the twin-rich region, while the inset table lists all $J_c$ values determined at 77 K using either MOI (M) or direct transport (T) method (see text for definition).

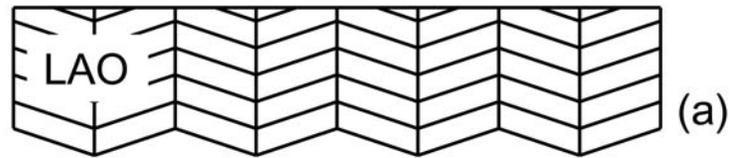

(a)

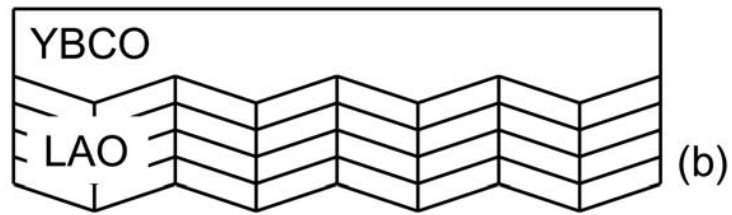

(b)

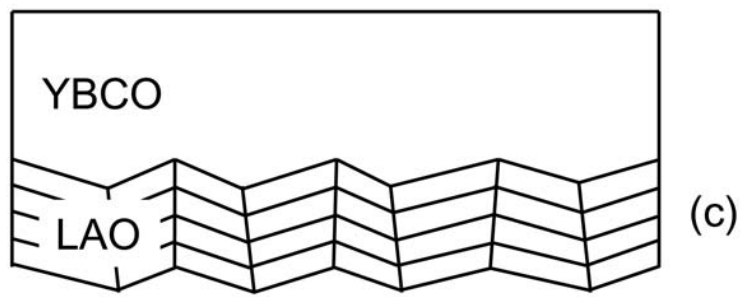

(c)

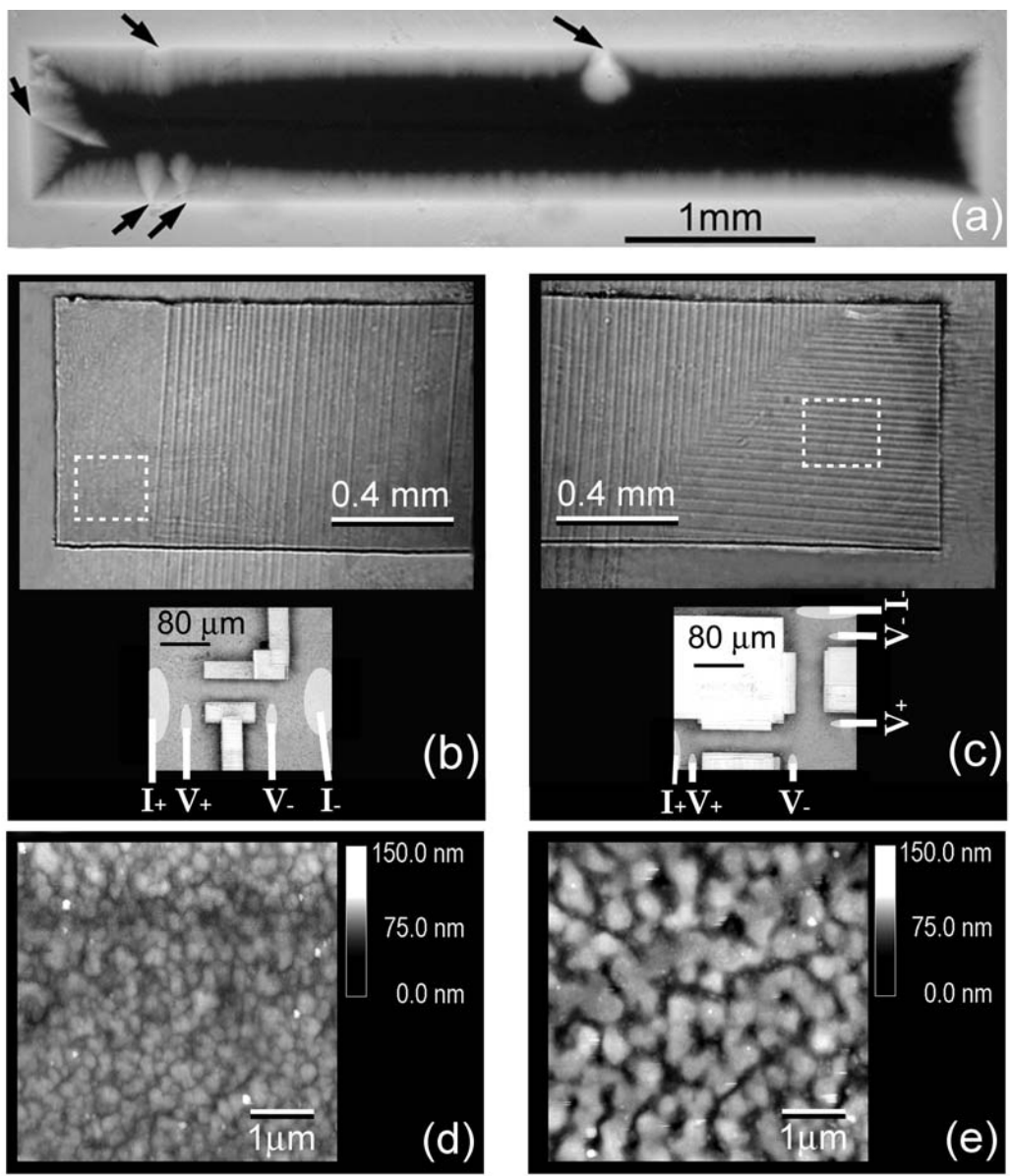

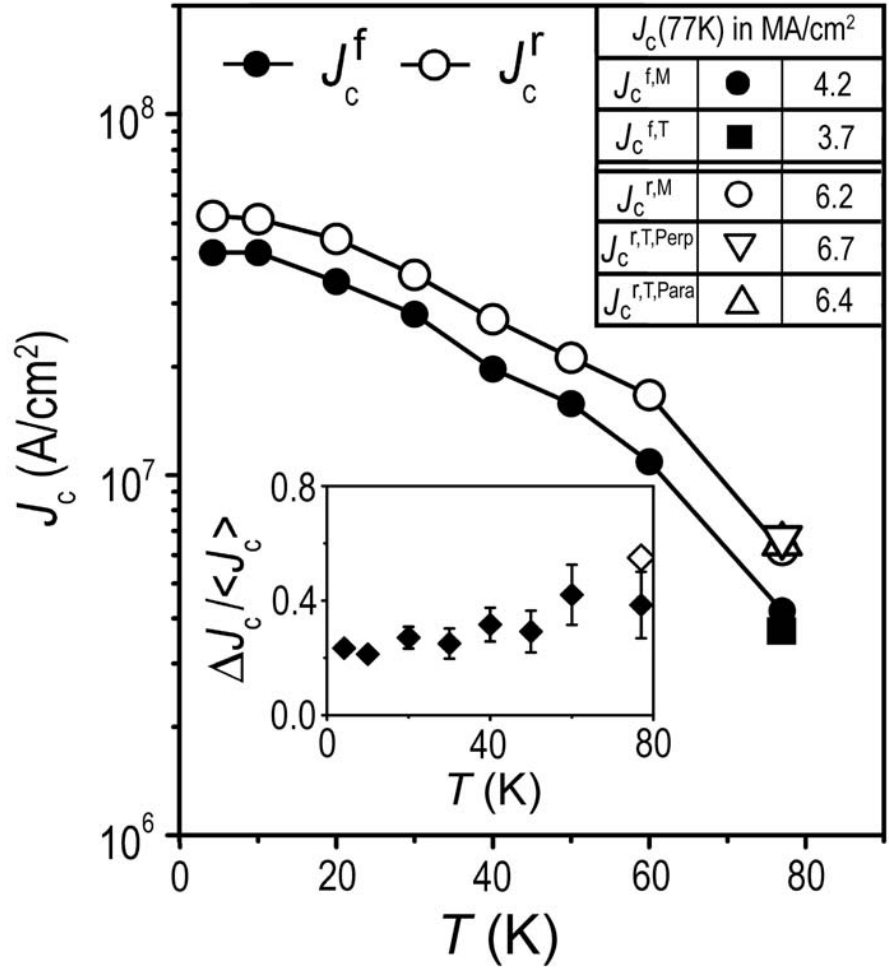